\documentstyle[multicol,aps,prl,epsfig]{revtex}

\begin{document}
\draft
\title{Pressure dependence of the spin gap in BaVS$_3$}
\wideabs{
\author{I. K\'{e}zsm\'{a}rki$^{1}$, Sz. Csonka$^{1}$, H. Berger$^{2}$,
L. Forr\'o$^{2}$, P. Fazekas$^{3}$, and G. Mih\'aly$^{1,2}$}

\address{$^{1}$Department of Physics, Institute of Physics, Budapest University
of Technology and Economics\\H-1111 Budapest, Hungary \\
$^{2}$Department of Physics, Ecole Polytechnique Federale de Lausanne\\
CH-1015 Lausanne, Switzerland\\
$^{3}$Research Institute for Solid State Physics and Optics\\
H-1525 Budapest, P.O.Box 49, Hungary \\}
\date{\today}
\maketitle

\begin{abstract}
We carried out magnetotransport experiments under hydrostatic pressure
in order to study the nature of the metal-insulator transition in BaVS$_3$.
Scaling relations for $\rho(T,H,p)$ are established and the pressure
dependence of the spin gap is determined. Our new results, in conjunction with
a re-analysis of earlier specific heat and susceptibility data, demonstrate
that the transition is weakly second order. The nature of the phase diagram in
the $T$--$p$--$H$ space is discussed.
\end{abstract}

}

There has been great progress in understanding metal-to-insulator transitions
which are accompanied by magnetic ordering and/or structural distortions
\cite{Imada}. In contrast, ``pure'' Mott transitions which do not involve any
apparent change of symmetry, remain little understood. The suppression of
magnetic order is usually due to frustration, or possibly orbital
fluctuations. Differences in lattice structure and the subspace of the
relevant orbitals make each of these systems unique.

The recent availability of single crystal specimens of BaVS$_3$ has brought
fresh insights \cite{Mihaly,Forro}, but understanding the character of its
Mott transition, and the nature of the low-temperature phases, remains a
challenging problem. Under atmospheric pressure, the material undergoes a
metal-to-insulator transition at $T_{\rm MI}=69{\rm K}$ from a $T>T_{\rm MI}$
bad metallic state to a $T<T_{\rm MI}$ insulator. There is a further
phase transition at $T_X=30{\rm K}$ \cite{Mihaly,Naka97}, and it is now
understood that long-period magnetic order sets in here \cite{RIKEN}. However,
in this paper we will be concerned with the intermediate ($T_X<T<T_{\rm MI}$)
insulating phase (we will call this the insulating phase) which is
isostructural with the metallic phase, and has no magnetic long range order
\cite{Sayetat,Ghedira}. It is a singlet insulator\cite{singlet} characterized
by a singlet--triplet gap. Various measurements give varying estimates for the
spin gap: NMR/NQR \cite{Naka97} suggested $\Delta_S=22{\rm meV}$,
but according to inelastic neutron scattering studies
$\Delta_S\sim 10{\rm meV}$ is more likely \cite{Naka99}. In any case, it is
remarkable that a spin gap $\Delta_S>k_BT$ appears in a 3D system, without the
formation of any static pattern of bonds.

Our paper has two main points. First, we use magnetoresistivity measurements
to determine the pressure dependence of the spin gap up to $p=15{\rm kbar}$.
Here we closely follow Booth et al. \cite{Booth} who did a similar
analysis for $p=1$bar. However, we managed to improve their procedure
by dropping some of the simplifying assumptions, and derive corresponding
statements from the data instead. -- Second, we
use our fresh data in conjunction with previous susceptibility \cite{Mihaly}
and specific heat \cite{Imai} data to discuss the properties of the
thermodynamic potential $G(T,p,H)$, and determine the character of the
metal--insulator transition. In contrast to previous claims \cite{Graf}, we
conclude that it is a genuine second order phase transition.

Single crystals of BaVS$_3$ were grown by Tellurium flux method. The
samples used in this work come from the same batch as those
investigated in Ref. \cite{Mihaly}. The resistivity measurements
under pressure were performed in a non-magnetic copper--beryllium cell
using kerosene as pressure
medium. The magnetoresistance (MR) was studied in two ways: by magnetic
field sweeps at various temperatures and by measuring the temperature
dependence of the resistivity in zero and in high magnetic field
($H_{\rm max}=120$kG). By simultaneous application of carbon glass and
capacitive thermometers the overall uncertainty of temperature was
reduced below $\pm0.05$K.

\begin{figure}
\noindent
\centerline{\includegraphics[width=0.7\linewidth]{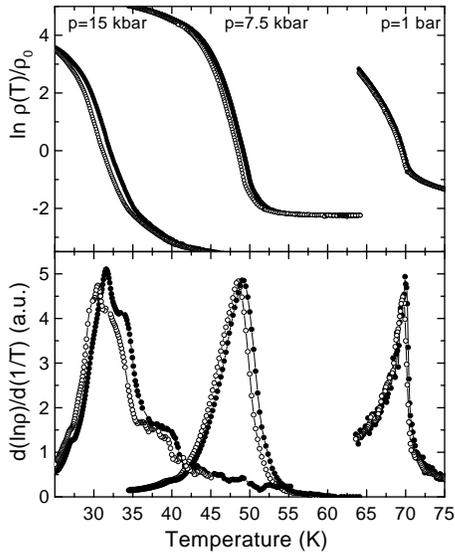}}
\caption{\it Temperature dependence of the normalized resistivity
(upper panel) and its logarithmic derivative (lower panel) at
various pressures. The sharp peaks define the transition
temperatures at $p=1$bar, 7.5kbar, and 15kbar.} \label{fig:MR1}
\end{figure}

Figure \ref{fig:MR1} shows representative $\rho(T)$ curves measured at
various pressures both in $H=0$ and $H=120{\rm kG}$ magnetic field.
In accordance with earlier observations \cite{Forro,Graf}, the dominant
effect is the
pressure induced reduction of the transition temperature. The influence of
the applied magnetic field is comparatively weak. It is an
excellent approximation that for all pressures the resistivity
depends on the magnetic field only through the field-induced shift
of $T_{\rm MI}$; the resistivity in $H_{\rm max}$
is the shifted counterpart of the zero-field one. It seems a plausible
assumption that any smaller field has the same effect on the resistivity,
i.e.,
\begin{equation}
\rho (T,H,p)=\rho\left((T-T_{\rm MI}(H,p),p\right).
\label{eq:1}
\end{equation}
$T_{\rm MI}$ itself can be determined with a high accuracy by the logarithmic
derivative of the resistivity, $\partial (ln\rho )/\partial (1/T)$, as
shown in the lower panel of Fig. \ref{fig:MR1}.

The field dependence of the transition temperature is weak, but it is
increasing with increasing hydrostatic pressure:
$\Delta T_{\rm MI}(H_{\rm max})$ grows from $0.35$K at ambient pressure
to $0.9$K at 15kbar.

\begin{figure}
\noindent
\centerline{\includegraphics[width=0.8\linewidth]{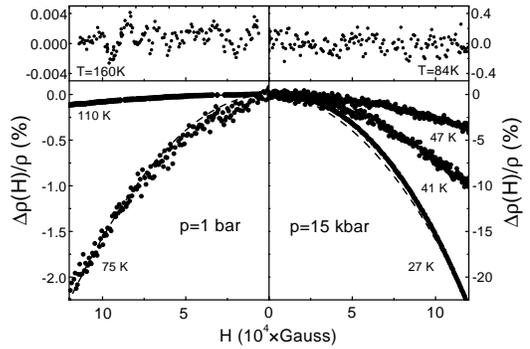}}
\caption{\it Magnetic field dependence of the resistivity at
various temperatures at $p=1$bar, and $p=15$kbar. The dashed
curves indicate the quadratic fit to the data. Note the different
scales in the upper panel.} \label{fig:hsq}
\end{figure}

As expected from the data shown in Fig.~\ref{fig:MR1},  the magnetoresistance
becomes large in the vicinity of the metal--insulator transition.
Field sweeps performed at constant temperatures revealed a quadratic
magnetoresistance for each pressure,
\begin{equation}
\Delta \rho (T,H,p) \propto H^2 .
\label{eq:2}
\end{equation}
Figure \ref{fig:hsq}  shows representative curves for $p=1$bar,
and $p=15$kbar. Note that far above the transition the
magnetoresistance is strongly reduced (upper panels). In fact, for
$T>2T_{\rm MI}$, no effect can be observed, confirming that the
dominant term in the magnetoresistance is due to the field
dependence of $T_{\rm MI}$ and any other contribution to $\Delta
\rho$ is below the detection limit.

Taken in conjunction with Eqn. (\ref{eq:1}), Eqn. (\ref{eq:2}) shows
that at the field strengths available to us, $\Delta T_{\rm MI}(H)$ is
purely quadratic:
\begin{equation}
\Delta T_{\rm MI}(p,H) = -a(p) H^2\, .
\label{eq:3}
\end{equation}

In the sense of Eqn. (\ref{eq:1}), at any given pressure $\Delta\rho(H)$ can
be expanded in terms in $\Delta T_{\rm MI}(H)$, or effectively in
powers of $H$. Up to
second order in $\Delta T_{\rm MI}$
\begin{eqnarray}
&& \Delta \rho (T,p,H)  =  \left.-\frac{\partial
\rho(T,p)}{\partial T}\right|_{H=0} \Delta T_{\rm MI}(p,H)
+\nonumber \\[2mm] && \frac{1}{2} \left.\frac{\partial^2
\rho(T,p)} {\partial T^2}\right|_{H=0}\Delta T_{\rm MI}(p,H)^2 =
\label{eq:expa}\\[2mm] && a(p) \left.\frac{\partial
\rho(T,p)}{\partial T}\right|_{H=0}H^2 + \frac{1}{2}a^2(p)
\left.\frac{\partial^2 \rho(T,p)}{\partial T^2}\right|_{H=0}
 H^4.
\label{eq:expb}
\end{eqnarray}
Though  $\Delta T_{\rm MI}(H)$ contains no fourth-order shift, the
$\propto H^4$ term appearing in Eqn. (\ref{eq:expb}) is not negligible because
$\rho(T)$ is a sharply varying function of $T$ near $T_{\rm MI}$.

\begin{figure}[!t]
\noindent
\centerline{\includegraphics[width=0.8\linewidth]{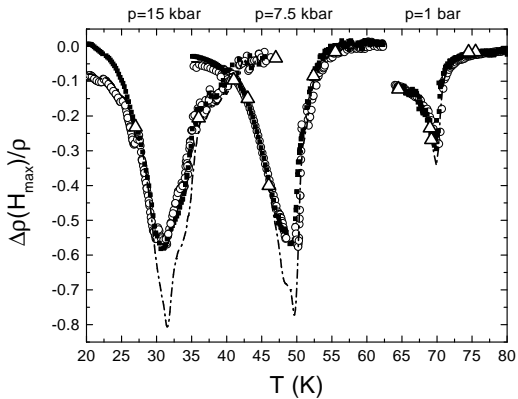}}
\caption{\it Analysis of the magnetoresistance at various
pressures. Full symbols: $\Delta \rho (T,H_{\rm max})/\rho (T)$
measured by thermal sweeps in presence of magnetic field,
triangles: results of measurements by magnetic field sweeps at
selected temperatures, dashed lines: first order calculation
according to Eqn. (\ref{eq:expa}); open circles: second order
calculation from the zero field $\rho(T)$ curves according to Eqn.
(\ref{eq:expa}).} \label{fig:derho}
\end{figure}

The consistency of our scheme can be verified, since the quantities on either
side of Eqn. (\ref{eq:expa}) can be measured
independently. In Fig.~\ref{fig:derho} full squares show the directly measured
$\Delta \rho(T,H_{\rm max})/\rho (T)$,  as obtained from the difference
of the $\rho(T)$ curves measured in 120kG, and in zero-field, resp.
This set of data is to be compared to the curves calculated
from (\ref{eq:expa}) using the partial derivatives of resistivities
measured at zero field, and inserting the value of
$\Delta T_{\rm MI}(H_{\rm max})$ determined experimentally
(Fig.~\ref{fig:MR1}, upper panel). There is no free parameter in the
calculation. The results of expanding to order $(\Delta T_{\rm MI})^2$
are shown in Fig.~\ref{fig:derho} by open circles. The agreement is good.
It is clear that the second term of the expansion is necessary to describe
the results near $T_{\rm MI}$, and that it is also sufficient.

The first order expansion (shown by dashed line in Fig.~\ref{fig:derho})
gives already a good approximation except in the closest vicinity of the
transition. In the range where the first order approximation is valid,
\begin{equation}
\Delta \rho (H,p)= \left. \frac {\partial \rho (T,p)} {\partial
T}\right|_{H=0} \Delta T_{\rm MI}(H,p)=-\beta(p)H^2 \label{eq:5}
\end{equation}
i.e., the magnetoresistance curves determined by field sweeps
scale to the magnetic field dependence of the transition
temperature. The magnetoresistance has been determined at several
temperatures at each pressure; the corresponding values at $H_{\rm
max}$ are shown in Fig.~\ref{fig:derho} by open triangles. This
reasserts that the field dependence of the transition temperature
$T_{\rm MI}(H)$ is quadratic in $H$ for all pressures.

Eqn. (\ref{eq:3}) can be brought to the dimensionless form
 \begin{equation}
\frac{\Delta T_{\rm MI}(H)}{T_{\rm MI}}=-
\gamma \left(\frac{gS\mu_BH}{k_BT_{\rm MI}}\right)^2,
\label{eq:gamma}
\end{equation}
where $S=1/2$ and $g=2$. The sensitivity of the transition to the applied field is
characterized by the constant $\gamma$. The magnetoresistance results yield
$\gamma =0.45$, {\sl independently of pressure}.

We note that an analogous relationship describes the suppression
of the spin-Peierls transition by a magnetic field, with very
similar values of $\gamma$.  On the theoretical side, Bulaevskii
et al. \cite{Bulaevskii} predicted $\gamma =0.44$, while Cross
\cite{Cross} found $\gamma =0.38$. Later experiments established
that $\gamma =0.41\pm 0.05$ is universal both for
organic\cite{Northby} and inorganic \cite{Hase} compounds. We may
hope useful insights from such parallels, though the spin-Peierls
transition involves a static distortion which is not the case with
BaVS$_3$. Since the magnetic field acts by making an antiparallel
alignment of magnetic moments less favourable, we guess that the
nature of short-range correlations in BaVS$_3$ is similar as in
spin-Peierls systems.

Another dimensionless form of (\ref{eq:3}) is obtained by introducing the
pressure dependent spin gap $\Delta_S(p)$:
\begin{equation}
\frac{\Delta T_{\rm MI}(H)}{T_{\rm MI}}=
-\alpha\left(\frac{g\mu_BSH}{\Delta_S(p)}\right)^2
\label{eq:alpha}
\end{equation}
assuming that it is related to a characteristic magnetic field $H_c$ via the
corresponding Zeeman energy. We derived the coefficient $\alpha=1.0\pm 0.1$ by
accepting  the ambient pressure value of the spin gap
$\Delta_S(p=1{\rm atm})=10{\rm meV}$ \cite{Naka99}.  $H_c$ can be also
understood as the critical magnetic field which would suppress the
metal-to-insulator transition completely but one has to bear in mind that only
the weak-field behavior is known (with our assumption $H_c=1700{\rm kG}$).
Analyzing their $p=1{\rm atm}$
magnetoresistivity data, Booth et al. found $H_c=2600{\rm kG}$ (175K)
\cite{Booth}. Let us note, though, that Booth et al. postulated a specific
(elliptical) form of the phase boundary in the $T$--$H$ plane, and thus in
their scheme, $\alpha$ is fixed by geometry rather than measured.

\begin{figure}
\noindent
\centerline{\includegraphics[width=0.7\linewidth]{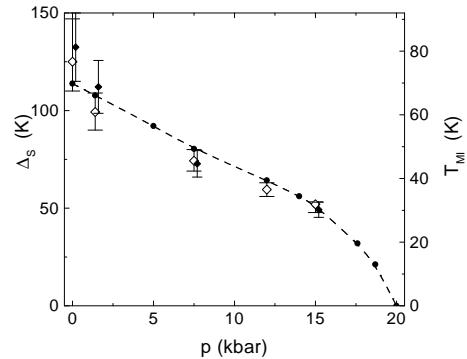}}
\caption{\it The pressure dependence of the spin gap (diamonds,
left-hand scale) scales with that of $T_{\rm MI}$ (circles,
right-hand scale). The spin gap determined from the shift of the
transition temperature and from the magnetoresistance is shown by
open and full diamonds, respectively.} \label{fig:bound}
\end{figure}

Equation (\ref{eq:alpha}) allows to determine the pressure
dependence the spin gap $\Delta_S(p)$ directly from the measured
$T_{\rm MI}$ shifts (Fig.~\ref{fig:MR1}). They can also be derived
from magnetoresistance measurements performed at constant
temperatures using Eqs. (\ref{eq:5}) and (\ref{eq:alpha}) which
relate $\alpha$ to $\beta(p)$. The results are shown in
Figure~\ref{fig:bound} where the pressure dependence of the
transition temperature is also presented. We found that within the
experimental resolution the spin gap of the insulating state
scales with the transition temperature according to $\Delta_S(p)
\approx  1.7\cdot k_{\rm B}T_{\rm MI}(p)$.

The transition temperature $T_{\rm MI}$ is marked as a peak in the
$d(ln \rho )/d(1/T)$ vs $T$ plots but it has been long disputed
whether it belongs to a true thermodynamic transition. The early
suggestion of a weakly first order transition \cite{Sayetat} was
discarded in favour of a smooth conductor-to-insulator crossover
\cite{Graf}. This might have been either of the kind of the
supercritical behavior observed near the critical point of the
V$_{2-x}$Cr$_x$O$_3$ system \cite{Imada}, or like the behavior of
Kondo insulators \cite{Aeppli}. However, we argue against this
interpretation, and show that the transition is a genuine phase
transition. As to the first alternative, the nearness to a
critical-point situation should be an accidental feature and we
should expect that changing the parameters drives either towards a
first-order phase transition, or well away from the critical
point. However, the resistivity curves show that essentially the
same transition is taking place at all pressures up to $p_{\rm
cr}=20{\rm kbar}$ \cite{Forro,high}, and/or in accessible magnetic
fields. -- As to the second possibility, the crossover from the
low-temperature insulator to the high-temperature conductor in
Kondo lattices can be generically continuous, because the
insulating ground state does not violate Luttinger's theorem.
Actually, this possibility cannot be trivially refuted for
BaVS$_3$, because the unit cell contains two V sites, i.e., an
even number of $3d$ electrons. However, we explicitly show below
that there is a phase transition at $T_{\rm MI}$.

Unfortunately, at $p>1{\rm bar}$, only resistivity data are known,
thus the following discussion has to be restricted to the ambient
pressure case. The lowest-order quantities which clearly show
non-analytic behavior are the temperature derivatives of the
susceptibility components: $d\chi_c/dT$ and $d\chi_a/dT$ have
large jumps at $T_{\rm MI}$ \cite{Mihaly,anis}. If it were only
for these, $\partial\chi/\partial T=-(\partial^3 G/\partial
T\partial H^2)$ being a third derivative of the thermodynamic
potential $G(T,p,H)$, one might have suspected that the transition
is of third order. We should, however, seek to relate $\Delta
(\partial \chi /\partial T)=(\partial\chi_{\rm I}/\partial T)-
(\partial\chi_{\rm M}/\partial T)$ to other thermodynamic
quantities by general reasoning \cite{fisher}.

The thermodynamic potential $G(T,p,H)$ must be continuous across the
transition:  $f(T,p,H)=G_I(T,p,H)-G_M(T,p,H)\equiv 0$ along the
phase boundary given by (\ref{eq:3}): $T_{\rm MI}(H)=T_{\rm MI}^0-a_0 H^2$,
where $T_{\rm MI}^0=T_{\rm MI}(p=1{\rm bar})$, and $a_0=a(p=1{\rm bar})$.
Expanding $f$ to fourth order in $H$, and considering that the entropy and
the linear susceptibility are continuous across the transition,
and that neither of
the phases has spontaneous magnetization  $M_I(T,H=0)=M_M(T,H=0)=0$, we are
left with the relationship
\begin{equation}
\Delta\left(\frac{\partial \chi}{\partial T}\right)    =\frac{a_0}{T}\Delta C
+\frac{1}{12a_0}\Delta\chi^{(3)}\, .
\label{eq:ehr2}
\end{equation}
The discontinuity of $\partial\chi/\partial T$ has to be balanced by a combination of the
discontinuities of the specific heat $C$, and the non-linear susceptibility
$\chi^{(3)}$.

\begin{figure}[t]
\noindent
\centerline{\includegraphics[width=0.9\linewidth]{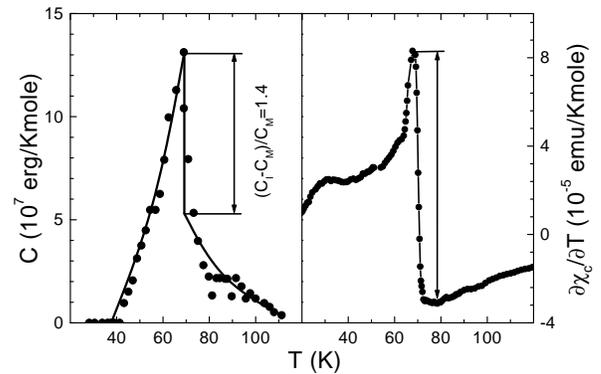}}
\caption{\it Comparison of the the specific heat (based on data
taken from \protect\cite{Imai}) and temperature derivative of the
$c$-axis susceptibility (based on \protect\cite{Mihaly}). The
lines are guide to the eye.} \label{fig:jumps}
\end{figure}

We have measured $\Delta(\partial\chi/\partial T)$ \cite{Mihaly},
and $a_0$. $\Delta C$ and $\Delta\chi^{(3)}$ have to be taken from
published data. Imai et al. \cite{Imai} find a strong anomaly at
$\sim$69K in the electronic specific heat which they chose to
identify as a sharp peak rather than as a discontinuity but this
interpretation is not compelling. We have made a new plot using
their published data points and find that they allow that part of
the peak height is made up by a discontinuity
(Fig.~\ref{fig:jumps}, left). Our fit with $(C_{\rm I}-C_{\rm
M})/C_{\rm M}=1.4$ is certainly somewhat arbitrary, but the
correct value cannot be very different.

Though the so obtained $\Delta{C}$ is of the order of magnitude
required by (\ref{eq:ehr2}), a fraction of
$\Delta(\partial\chi/\partial T)$ has to be matched by
$\Delta\chi^{(3)}$. High-field magnetization data have been
published by Booth et al. \cite{Booth}. The authors emphasize that
the magnetization curves are essentially linear, but in fact the
data shown in their Fig. 2 are compatible with a size of the
discontinuity $\Delta\chi^{(3)}<0$, which is sufficient to satisfy
(\ref{eq:ehr2}).

The character of the transition is described by Eqn. (\ref{eq:ehr2}) which
contains both second and third derivatives of $G$. For this reason, one
might say that the continuous phase transition is ``weakly second order''.
In any case, the existence of a surface of continuous phase transitions in the
$T$--$p$--$H$ space strongly suggests that there is a distinct non-magnetic
insulating phase which differs
from the metallic phase not in transport properties only, but also in the
sense of possessing a hidden order.

In conclusion, magnetotransport measurements under pressure were performed
on single crystals of BaVS$_3$ in order to study the nature of the
metal--insulator transition. We determined the pressure dependence of the
spin gap in the insulating phase and showed that $\Delta_S$ scales with
the transiton temperature. We discussed the nature of the phase diagram and
pointed out that the metal--insulator transition is not a smooth crossover
but a genuine phase transition.

{\bf Acknowledgements}. This work was supported by the Swiss National
Foundation for Scientific Research and by
Hungarian Research Funds OTKA T025505, FKFP 0355, and AKP 98-66.


\begin{references}
\bibitem{Imada}M. Imada et al., Rev. Mod. Phys. \textbf{70}, 1039 (1998).
\bibitem{Mihaly}G. Mih\'aly et al., Phys. Rev. B \textbf{61}, R7831 (2000)
\bibitem{Forro}L. Forr\'o et al., Phys. Rev. Lett. (August 28, 2000).
\bibitem{Naka97} H. Nakamura, H. Imai and M. Shiga, Phys. Rev. Lett.
{\bf 79}, 3779 (1997).
\bibitem{RIKEN} M. Shiga and H. Nakamura, RIKEN Review No. 27, 48 (2000).
\bibitem{Sayetat}F. Sayetat et al., J. Phys. C \textbf{15}, 1627 (1982)
\bibitem{Ghedira}M. Ghedira et al., J. Phys. C \textbf{19}, 6489 (1986)
\bibitem{singlet} The finding of the susceptibility anisotropy \cite{Mihaly}
shows that spin--orbit coupling is not negligible, and the notion of a
singlet is only approximate.
\bibitem{Naka99} H. Nakamura {\sl et al.}, J. Phys. Chem. Solids {\bf 60},
 1137 (1999).
\bibitem{Booth}C. H. Booth et al., Phys. Rev. B {\bf 60}, 14852 (1999).
\bibitem{Imai} H. Imai, H. Wada and M. Shiga, J. Phys. Soc. Jpn. {\bf 65},
3460 (1996).
\bibitem{Graf} T. Graf {\sl et al.}, Phys. Rev. B {\bf 51}, 2037 (1995).
\bibitem{Bulaevskii} L. N. Bulaevskii, A. I. Buzdin and D. I. Khomskii,
Solid State Commun.  {\bf 27}, 5 (1978).
\bibitem{Cross} M. C. Cross, Phys. Rev. B {\bf 20}, 4606 (1979).
\bibitem{Northby} J. A. Northby {\sl et al.}, Phys. Rev. B {\bf 25}, 3215
(1982) and references therein.

\bibitem{Hase} M. Hase, I. Terasaki and K. Uchinokura, Phys. Rev.
Lett. {\bf 70}, 3651 (1993).
\bibitem{Aeppli} G. Aeppli and Z. Fisk: Comments Cond. Mat. Phys. {\bf 16},
  155 (1992).
\bibitem{high} Actually, it is possible that at the highest pressures,
there is a metal-to-magnetic-insulator transition. We have not seen any sign
of this up to $p=15{\rm kbar}$.
\bibitem{anis} The difference between the $d\chi_c/dT$ and $d\chi_a/dT$
discontinuities being of the order of 10\%, we neglect anisotropy in the
discussion.
\bibitem{fisher} For Ising antiferromagnets, the Fisher relation (M.E.
Fisher: Phil. Mag. {\bf 7}, 1731 (1962)) $a_0 C=d(\chi T)/dT$ has long
been used to establish
a relationship between the divergence of $C$, and the infinite slope of $\chi$
at $T_N$. The relationship holds for strongly uniaxial systems, but
would not apply to nearly isotropic systems like BaVS$_3$. It is nevertheless
interesting that if we assumed that
$d\chi/dT$ has a finite discontinuity, then also Fisher's equation would
say it is related to a jump of $\Delta C$.






\end{references}
\end{document}